\documentclass[prl,preprint,showpacs,preprintnumbers,amsmath,amssymb,superscriptaddress]{revtex4}

\usepackage{graphicx}
\usepackage{dcolumn}
\usepackage{bm}

\usepackage{graphicx}
\usepackage{psfrag}

\newcommand{\R}{{\mathbb R}}

\def\hat{\widehat}
\def\tilde{\widetilde}
\def \bfo {\begin {eqnarray*} }
\def \efo {\end {eqnarray*} }
\def \ba {\begin {eqnarray*} }
\def \ea {\end {eqnarray*} }
\def \beq {\begin {equation}}
\def \eeq {\end {equation}}

\def \det {\hbox{det}}

\def \e {\varepsilon}
\def \p {\partial}

\def\p{\partial}

\def\R{\mathbb R}

\def\mh{\hat{m}}
\def\rx{{\mathbf r}}
\def\ry{\tilde{\mathbf r}}

\def\eo{E_j^{\pm,1}}
\def\er{E_j^{\pm,R}}

\begin{document}

\preprint{}

\title{Approximate quantum cloaking and almost trapped states}

\author{Allan Greenleaf\,${}^*$}
\affiliation{Department of Mathematics
University of Rochester, Rochester, NY 14627}

\author{Yaroslav Kurylev}
\affiliation{Department of Mathematical Sciences, University College London, Gower Str, London,
WC1E 6BT, UK}

\author{Matti Lassas}
\affiliation{Institute of Mathematics, Helsinki University of Technology, FIN-02015, Finland}

\author{Gunther Uhlmann}
\affiliation{$\hbox{Department of Mathematics, University of Washington, Seattle, WA 98195}$\\
${}^*$Authors listed in alphabetical order}


\begin{abstract}

We describe families of potentials which act as approximate cloaks for matter waves, i.e.,
for solutions of the time-independent Schr\"odinger equation at energy $E$, with
applications to the design of ion traps. 
These are derived from perfect cloaks for the conductivity and Helmholtz equations, by a procedure
 we refer to as  isotropic transformation optics.
If $W$ is a potential which is surrounded by a sequence $\{V_n^E\}_{n=1}^\infty$ of  
approximate cloaks, then for 
generic $E$, asymptotically  in $n$ (i) $W$ is both undetectable and unaltered by matter waves
 originating externally to the cloak; and (ii) the combined potential $W+V_n^E$ 
 does not perturb  waves outside the cloak. On the other hand,
for $E$ near a discrete set of energies, cloaking {\it per se} fails and the approximate
cloaks support wave functions concentrated, or {\it almost trapped}, inside the cloaked region and negligible
outside.
Applications  include  ion  traps,
almost invisible to matter waves or customizable to support
almost trapped states of arbitrary multiplicity. Possible uses include simulation of abstract quantum systems, magnetically tunable quantum beam switches, and
illusions of singular magnetic fields .

\end{abstract}


\maketitle

\emph{Introduction.}  A fundamental problem  is to describe the scattering of waves
at  energy $E$  by
a potential $V(x)$, as governed by the time-independent Schr\"odinger equation;
the related
inverse problem  consists of unique determination of $V$ from
scattering data or
from boundary measurements. 
In this Letter,  for a  generic energy $E$, we construct  families
$\{V_n^E\}_{n=1}^\infty$ of bounded potentials which 
 act as an \emph{approximate
quantum cloaks}:
for any potential $W$ whose support is surrounded by the support of $V^E_n$,  the scattering amplitude
of $W+V_n^E$ goes to zero as $n \to\infty$, so that $W$ is,
 asymptotically in $n$,  undetectable by waves at energy $E$. 
The central potentials $V_n^E$ are supported in a spherical annulus $\{r_1\leq r\leq r_2\}$ in $\R^3$ and  
have spatial oscillations of increasing amplitude as $r\searrow r_1$ 
and decreasing quasiperiod as $n\to\infty$,
so that  their local (resp., long range) effect on wave propagation  becomes stronger 
(resp., weaker) as $n\to\infty$.
For generic $E$, the potential $W+V_n^E$ has a negligible effect on matter
 waves originating outside of its support.
Alternatively, for $E$ close to special values, $V_n^E$ allows the core $\{r\leq r_1\}$  to support
 {\it almost trapped states} and be used to form  traps for ions (here  denoting  any charged particles),
 almost invisible to external matter waves.  
An approximate version of the dichotomy regarding ideal cloaks for the Helmholtz equation
\cite[Thm.\ 1]{GKLU1} holds: If  $E$ is sufficiently separated from all eigenvalues, then with high 
 probability the approximate cloak   keeps particles of energy $E$ from entering the cloaked region, see Figs. 1(r,red) and  2(l); 
 on the other hand,
for $E$ close to an eigenvalue, the cloaked region supports almost trapped states, 
accepting and binding such particles, leading to a new type of ion trap, cf.
Figs. 1(r,blue) and 2(r) and formula (\ref{The new equation}).

Recently, Zhang, et al., \cite{Zhang}, using ideas from transformation optics \cite{WardPen,PSS},  described an ideal quantum
mechanical cloak
at any fixed energy $E$ and proposed a physical implementation. 
The construction  starts with a
homogeneous, isotropic mass
tensor $\mh_0$ and  potential $V_0\equiv 0$, and subjects this pair to the 
singular
transformation  (\ref{F}) below.
The resulting $\mh, V$ yield a Schr\"odinger equation that is in fact  the Helmholtz equation  for the corresponding singular Riemannian metric and thus
covered by the analysis of  cloaking for the Helmholtz equation   in 
\cite{GKLU1}. The treatment there
 shows
that  the potential within the cloaked region is
undetectable by
exterior measurements, and that the perfectly reflective Neumann boundary condition  
automatically holds for finite energy distributional solutions
at the inside of
the  cloaking
surface (the inner surface of the cloak),  $\Sigma$. 
The cloaking  mass tensor $\mh$ and potential are both singular, and $\mh$
infinitely anisotropic, at $\Sigma$. These combine to make such a quantum cloak challenging to construct, with ultracold atoms in an optical lattice proposed in \cite{Zhang} as a possible realization. 

The approach in this Letter  is to forgo the perfect functioning of the ideal quantum cloak, and to describe bounded potentials
which function as  approximate  cloaks with respect to the isotropic $\mh_0$, with
the failure to  cloak perfectly in fact  a controllable feature that can be taken advantage of for applications described below. The  $V_n^E$  are found 
by means of {\it isotropic transformation optics}, a technique we introduce for avoiding the singular and anisotropic behavior, difficult to physically realize, of material parameters that commonly occur in transformation optics-based designs \cite{PSS,Le,LePhil}; more details and proofs  can be found  in \cite{GKLU5}.

\emph{Inverse scattering and conductivity cloaks.}
 There is an enormous literature 
on unique determination of a potential $V$ from scattering of plane waves at energy $E$ by
 the Schr\"odinger equation $\left(-\nabla^2 + V\right)\psi=E\psi$, as encoded in the scattering
amplitude $a_V(E,\theta,\omega)$. Equivalently, for compactly supported $V$ one
may consider the near-field measurements of wave functions  at the boundary of a large
region $\Omega$, as encoded in the 
Dirichlet-to-Neumann (DN) operator, $\Lambda_V(E)(\psi|_{\p\Omega})=\p_{\nu}\psi|_{\p\Omega}$ 
where $\nu$ is the normal of $\p \Omega$ \cite{ADD2}. Ideal quantum cloaking can be considered as giving 
highly singular examples of nonuniqueness, but in order
to construct  approximately cloaking potentials $V_n^E$, 
we first need to recall the ideal cloaking conductivity $\sigma_1$ of \cite{GLU3}. 
For simplicity, we describe the cloak on $B_3-B_1$,
 with $B_R$ denoting the central ball of radius $R$ in $\R^3$, so that the {\it cloaking surface},
 the interface between the cloaked and uncloaked regions, is $\Sigma=\{r=1\}$.  
Subjecting a homogeneous, isotropic conductivity $\sigma_0$ to a singular transformation (``blowing up a point"), we
constructed \cite{GLU3} certain  singular, anisotropic conductivity tensor fields on $B_3- B_1$ which, 
when augmented by any nonsingular (bounded above and below) conductivity on
$B_1$, result in  conductivities on $B_3$ giving  the same electrostatic
boundary
measurements as  $\sigma_0$.
(For related results in dimension two, see \cite{LTU,KSVW}.)
The same
construction, applied to the electric permittivity and magnetic
permeability rather than the conductivity tensor, was
used 
\cite{PSS}  to propose an  electromagnetic (EM) cloak at nonzero frequency, and a microwave experimental
realization of a variant of this design reported in
\cite{Sc}. Ray-based cloaking for dimension two was proposed in \cite{Le},
while potentials transparent for rays are in \cite{HHLe}.

To be precise,
 let $F=(F^1,F^2,F^3):B_3-\{0\}\to B_3-B_1$ be the singular transformation, for $\rx\in\R^3$,
\beq \label{F}
\ry=F(\rx)=\rx, r>2;\, F(\rx)=\left(1+\frac{r}2\right)\frac{\rx}{r},\,
0<r\le 2,
\eeq
which results in the transformed version of $\sigma_0$ on $B_3-B_1$, augmented for simplicity
by $2\sigma_0$ on $B_1$,
\beq \label{ideal1}
\sigma_1=F_*\sigma_0,\, \rx \in B_3-B_1;
\quad
\sigma_1=2 \, \sigma_0,\,  \rx \in B_1.
\eeq
Here, $F_*$ denotes the change-of-variables action of  $F$  on tensors $\sigma=(\sigma^{jk})$, 
writing $\rx=(x^1,x^2,x^3)$,
\beq\label{eqn-transf law}\nonumber
(F_*\sigma)^{jk}(\ry)=\left.
\frac 1{\det [\frac {\p F^j}{\p x^k}]}
\sum_{p,q=1}^3 \frac {\p F^j}{\p x^p}
\,\frac {\p F^k}{\p x^q}  \sigma^{pq}\right|_{\rx=F^{-1}(\ry)}.
\eeq 
This $\sigma_1$  has a singularity at  $\Sigma$,
both in that one of the eigenvalues (corresponding to the radial direction) tends to $ 0$ as
$r\searrow 1$, and that there is a jump discontinuity across 
$\Sigma$, within which $\sigma_1$ is non-singular.
Aside from the radius of the outer ball and the factor
$2$ in the second formula of  (\ref{ideal1}),
$\sigma_1$ is the conductivity  introduced in \cite{
GLU3} and shown
to be indistinguishable from $\sigma_0$, {\it vis-a-vis} 
boundary
measurements at $\p B_3$ of electrostatic fields. 

Consider also the corresponding acoustic equation,
\beq \label{acoustic-eq}
\p_i \left(\sigma_1^{ij} \p_j u\right) +E a_1 u=0, \quad a_1= \left(\hbox{det}[\sigma_1^{ij}]\right)^{-1}.
\eeq
Then, using  boundary
measurements at $\p B_3$ of the acoustic waves of frequency $\sqrt{E}$, the pair $\sigma_1, \, a_1$ is indistinguishable,  
  from $\sigma_0, \, a_0\equiv1$ \cite{GKLU1,ChenChan,Cummer,GKLU4}.

 In the case of ideal cloaking, the waves $u$  within the cloaked region
have a simple description \cite[Thm.1]{GKLU1}: (I)  if $E$ is  {\it not} a
Neumann eigenvalue of $-\nabla^2+W$ in the cloaked region, then $u$ must vanish there;
  or (II) if $E$ {\it is} an eigenvalue, then $u$ can be an associated eigenfunction there, while possibly vanishing outside of the cloaking surface. 
 The Dirichlet eigenvalues and eigenfunctions of (\ref{acoustic-eq}) on $B_3$ can be separated into
$(E_j^{+,1},u^+_j)$ and $(E_j^{-,1},u^-_j)$ with $u^+_j$ and $u^-_j$ supported in $B_3-B_1$ and
 $ B_1$, resp. 

\emph{Approximate  cloaks and failure of cloaking.} 
For  $1<R\le 2$, let $\sigma_R$
be given by the same formulae as in 
 (\ref{ideal1}), but applied on $B_3-B_R$ and $B_R$, resp., and similarly
for $a_R$ in  (\ref{acoustic-eq}). 
 Observe that for each $R>1$,  $\sigma_R$ and $a_R$ are {\it nonsingular}; however, their lower (resp., upper) bounds go to $0$ (resp., $\infty$)
as $R \searrow 1$. (Similar truncations of EM cloaks have been studied in \cite{RYNQ,GKLU3,KSVW}.) 

When $\sigma_1$, $a_1$ are replaced by $\sigma_R$, $a_R$ there is a decomposition similar
to the one above for (\ref{acoustic-eq}), with eigenvalues and eigenfunctions 
$(E^{+,R}_j,v^+_j)$ and $(E^{-,R}_j,v^-_j)$  concentrating in $B_3-B_1$ and 
 $B_1$, resp., and
 $\er$ converges to $\eo$ as  $R\to 1$. 
The solution of the boundary value problem 
$\p_i \left(\sigma_R^{ij} \p_j v\right) +E a_R v=0$ on $B_3$, $v|_{\p B_3}=f$, has an
eigenfunction expansion
\beq\label{The new equation}
v(x)=\sum_{\pm}\sum_{j=1}^\infty \left(\int_{\p B_3} f\frac {\p v_j^\pm}{\p \nu}\,dS\right)
\frac{v_j^\pm(x)}{E-\er}.
\eeq

An approximate version of dichotomy (I)-(II) holds for approximate cloaks:
When $E$ is not equal to any $\eo$,
one can  show the DN operators for the $\sigma_R, a_R$ converge 
 to that for $\sigma_1, a_1$ as $R\to 1$; physically, this means that the boundary  measurements 
 of pressure and the normal component of the particle velocity
for the approximate cloaks tend to those for the ideal cloak, 
which are themselves the same as for $\sigma_0, a_0$.
However,  if $E$ is close to some $\er$, the corresponding term
in (\ref{The new equation}) may dominate the others, in which case the solution $v$,
having a large coefficient of $v_j^-$,
concentrates in $B_1$. Since $v_j^-$ cannot vanish identically in
$B_3-B_1$,  both the near-field measurements on the boundary $\p B_3$ 
and far-field patterns differ from those 
corresponding to $\sigma_0$, $a_0$.  This interior resonance corresponds to a sound wave almost trapped within the cloak.

\emph{Isotropic transformation optics.}
A well known phenomenon in effective medium theory is  that homogenization of isotropic material parameters may lead, in the small-scale limit, 
to anisotropic ones \cite{Milt}. We  exploit this, using ideas from \cite{Allaire,Cherka},
 to approximate the anisotropic, almost cloaking $\sigma_R$ by {\it isotropic}
 conductivities $\sigma_{R,\epsilon}$ 
 so that for $\e>0$ the  $\sigma_{R,\epsilon}, \, a_R$  also function as approximate 
 acoustic  cloaks \cite{GKLU5}.
 The $\sigma_{R,\epsilon}(\rx)$ are  spatially highly oscillating, with  values varying roughly
 between the  extreme eigenvalues of $\sigma_{R}$ near $\rx$. 
 (In the context of EM cloaking, thin concentric layers of homogeneous, isotropic media were considered in \cite{HFJ,ChenChan2}.)

\emph{Approximate Schr\"odinger cloaks.}
The gauge transformation $\psi= {\sqrt \sigma} u$ reduces the acoustic  equation 
(\ref{acoustic-eq}), with nonsingular 
 isotropic conductivity $\sigma= \sigma_{R,\epsilon}$  in place of the anisotropic $\sigma_1$ and 
 $a_R$ in place of $a_1$, to the 
 Schr\"odinger equation
at the same energy $E$,
$\left(-\nabla^2+V^E_{R,\epsilon}\right)\psi=E \psi$,
where
$V^E_{R,\epsilon} = \nabla^2 ({\sqrt \sigma_{R, \epsilon}})/ {\sqrt \sigma_{R,\epsilon}}+E \left(1 - a_R^{1/2}\sigma_{R,\epsilon}^{-1} \right)$.
As $\sigma_{R,\epsilon}$ is highly oscillatory, $V^E_{R,\epsilon}$ consists of 
a quasiperiodic pattern of concentric radial potential barriers and wells of 
increasing amplitudes and decreasing widths as $\e\searrow 0$.
The radial profile of the potential over one quasiperiod (i.e.\ in 
one spherical layer) is in Fig.\ 1(left).
Since $\sigma_{R,\epsilon}=1$ near $\p B_3$, the boundary measurements of solutions of these 
Schr\"odinger equations at $\p B_3$ coincide with those for the 
corresponding acoustic equations.
By  the convergence for the acoustic equations, we can 
choose $R\searrow 1,\, \epsilon \searrow 0$, so that the 
boundary measurements for these 
Schr\"odinger equations converge to those for the acoustic
equation (\ref{acoustic-eq}) at energy $E$, which are the same as for  the Schr\"odinger equation in free space. Our main result is:
\smallskip

\noindent{\bf Approximate Quantum Cloaking.}
{\it Let $W $ be a bounded potential on $B_1$, and $E \in \R$ be neither a Dirichlet eigenvalue of the
free Hamiltonian $-\nabla^2$ on $B_3$ nor a Neumann eigenvalue of
$-\nabla^2+W$ on $B_1$.  Then   there exists a sequence of
cloaking potentials $V_n^E$ on $B_3$ such that
the DN operators
$\Lambda_{W+V_n^E }(E)  \rightarrow  \Lambda_0(E) $ as $n \to \infty.$
I.e., as $n\to\infty$ the potential $W+V_n^E $ is indistinguishable
at energy $E$
from the zero potential by near-field measurements; a similar result holds for far-field patterns. $W$ is thus approximately cloaked when surrounded by $V_n^E$.}

As any specific measurement device has a
limited precision,  this means that  it is possible to design a potential  to  cloak an object  within from
any single-particle measurements made using that device at energy $E$.
\smallskip

{\bf Numerics:} We use
analytic expressions to compute the wave function $\psi$
for an incident  plane wave with $\psi_{inc}(x)=ae^{ik\rx\cdotp \vec d}$. 
The computations are made without reference to physical units,
using
$a=1,E=0.5,k=\sqrt{E}$.
The cloak is based on $R=1.005$, corresponding to an anisotropy ratio of $\sigma_R$ at   
$\Sigma_R=\{r=R\}$ of $4\times 10^4$. 
 In the simulations we use a cloak consisting of 20 homogenized
layers inside and 30 homogenized layers outside of the cloaking surface
$\Sigma_R$.
Inside the cloak we have located a centrally symmetric step potential,
$W(x)=c_{inn}\chi_{[0,0.9]}(r)$. 
In the simulations, the cloaking potential $V_n^E$ and the energy $E$ are the same in all figures,
but we vary the constant $c_{inn}$.
In Fig.\ 1(r, red) and Fig.\ 2(l) we have 
$c_{inn}=-98.5$, and $\psi$ is
the wave in $\R^3$ produced by
 an incoming plane wave.
In Fig.\ 1(r, blue), with
$c_{inn}=+1.858$, and
in Fig.\ 2(r), with $c_{inn}=-71.45$, there is no incoming wave,
but an almost trapped state 
corresponding to an excited state
in the cloaked region.

\begin{figure}[htbp]
\vspace{-2.8cm}
\begin{center}
\centerline{
\includegraphics[width=.9\linewidth]{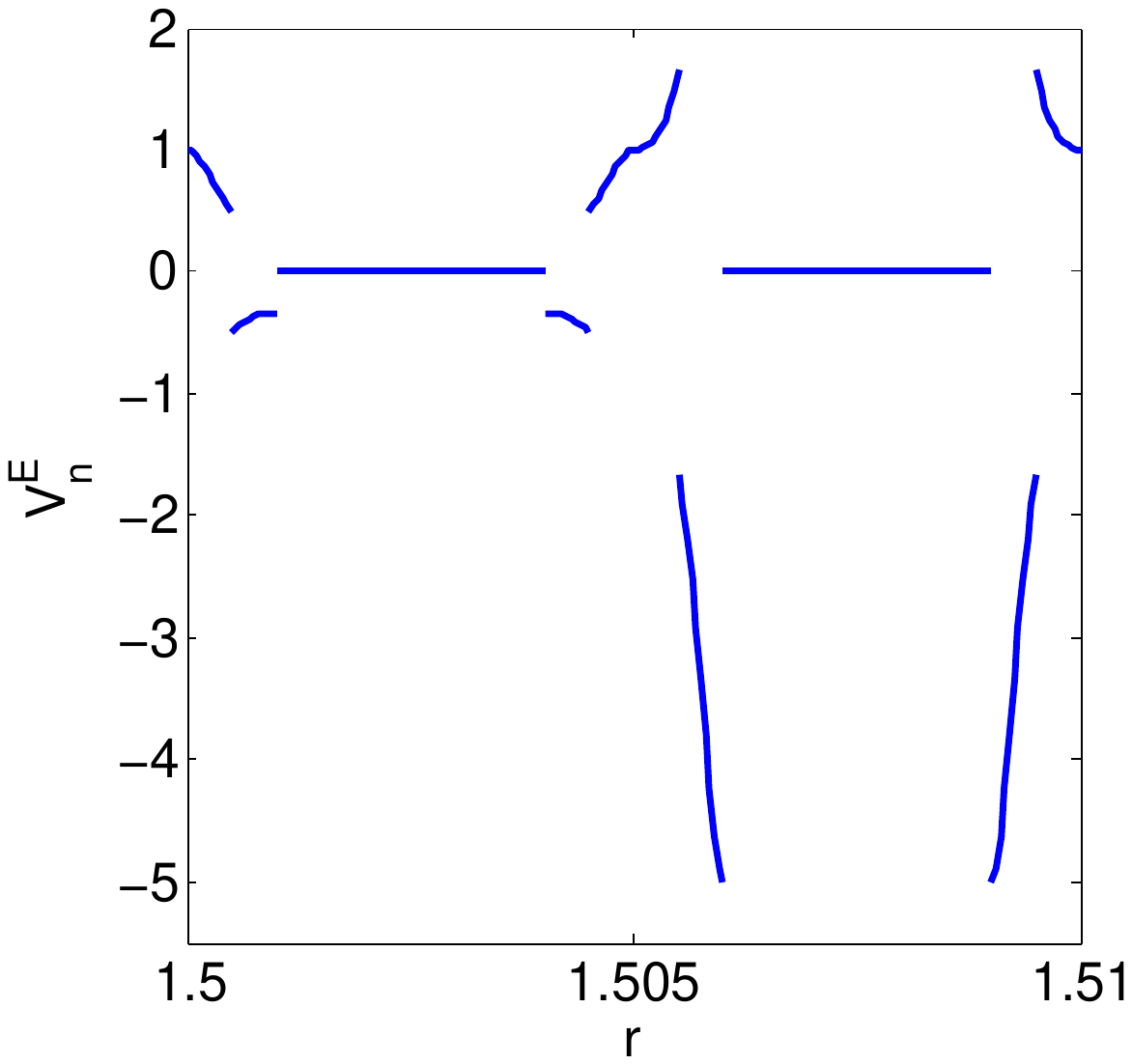}
\hspace{-3.4cm}
\includegraphics[width=.9\linewidth]{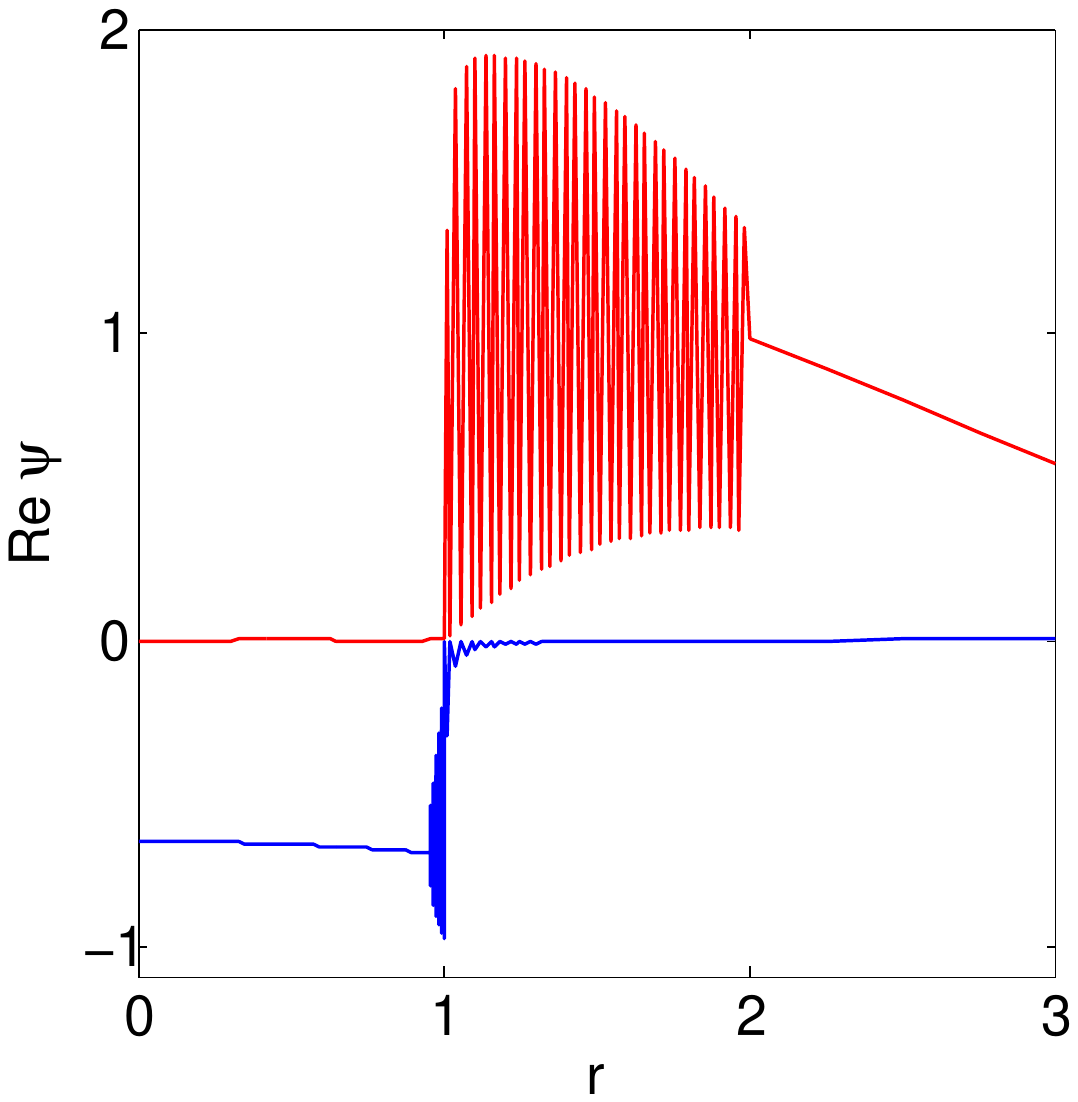}
}
\end{center}
\vspace{-3.5cm}
\caption{
{\bf Left:}
The radial profile of the potential $V_n^E$  over a typical  quasi-period $1.5<r<1.51$.
The potential $V_n^E$ in $R=1.005<r<2$  is 
obtained by repeating similar profiles, with increasing amplitudes as $r\searrow R$.
{\bf Right:}  $\hbox{Re}\, \psi$ on a segment
$\{(x,0,0):\ 0\leq x\leq 3\}$ for the
same cloaking potential $V_n^E$ and two different cloaked $W$'s. For the red curve,
$E$ is {\it not} close to an interior eigenvalue and $\psi$ is produced
by an incoming plane wave. For the blue curve, $E$ {\it is} a Dirichlet eigenvalue
on $B_3$ and $\psi$ is an almost  trapped state. 
}
\end{figure}

\begin{figure}[htbp]
\vspace{-2.3cm}
\begin{center}
\centerline{
\includegraphics[width=.75\linewidth]{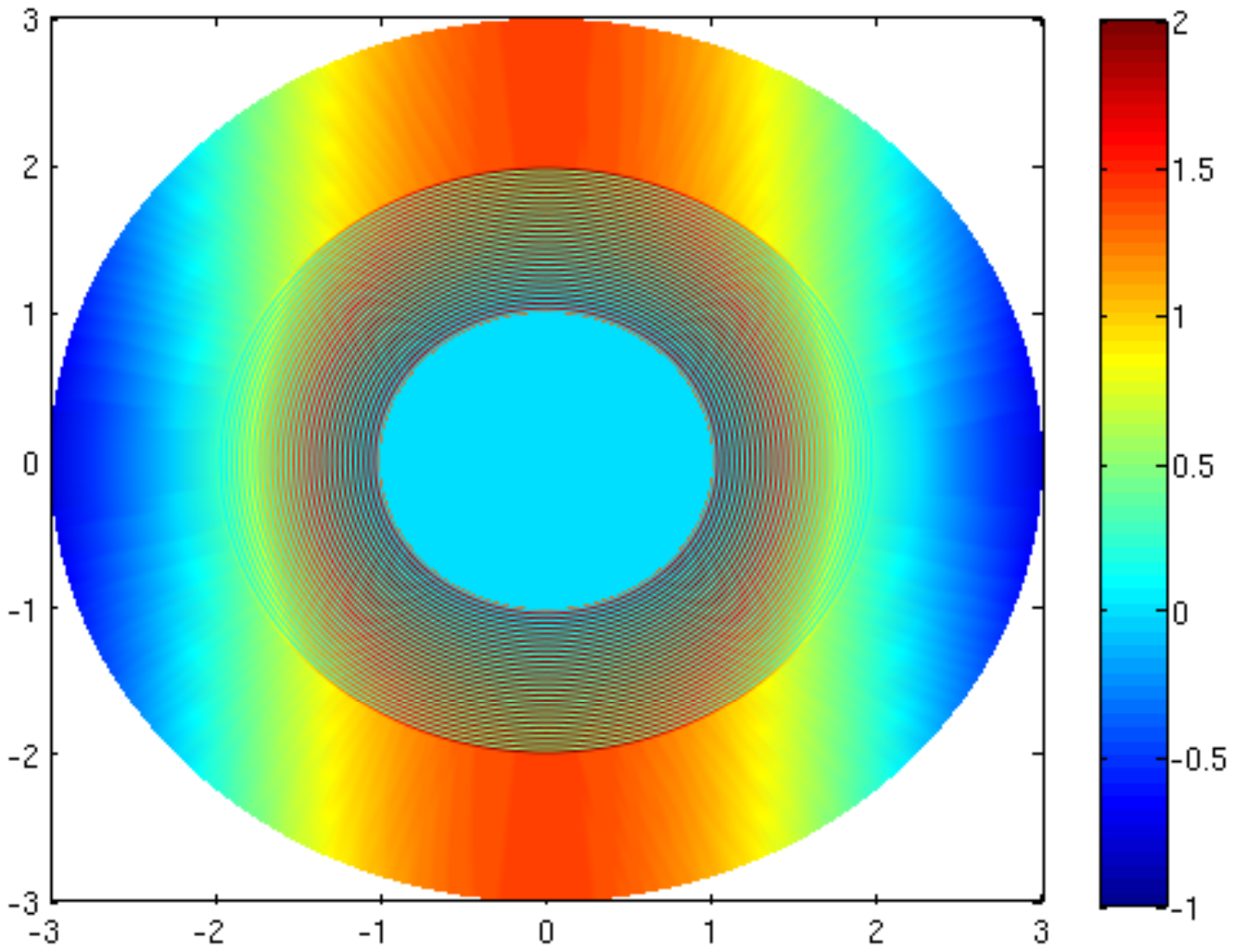}
\hspace{-2.1cm}
\includegraphics[width=.75\linewidth]{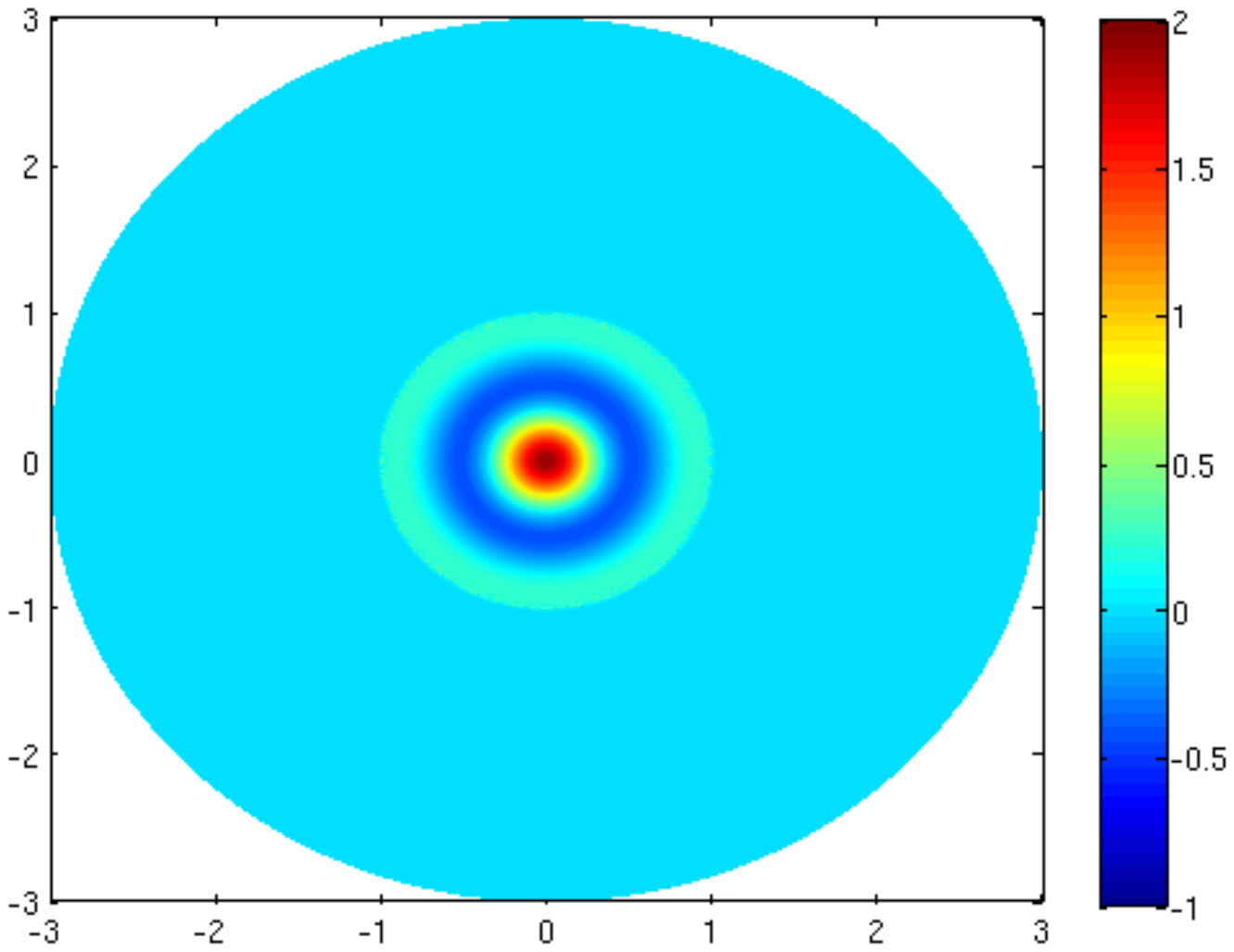}
}
\vspace{-3.7cm}
\end{center}
\vspace{-.2cm}

\caption{ 
{\bf Left:}  $\hbox{Re}\, \psi$ in $B_3$ for $\psi$ resulting from an incident plane wave.
$E$ is not near an interior Neumann eigenvalue; the matter wave passes unaltered. {\bf Right:} 
An  excited almost trapped state. $E=E_j^{-,R}$ is an energy
close to a Neumann
eigenvalue of  $B_1$,
for which the ideal cloak supports an   trapped state.}
\end{figure}

\smallskip
{\bf Applications:}\emph{ Almost trapped states and ion traps.}
A version of the dichotomy for approximate acoustic cloaks 
described above also  holds for approximate quantum cloaks, since the solutions differ only by a gauge transformation.
As a consequence, 
given an energy  $E$, the approximate cloak may be such  that  $E$ either is or is not an eigenvalue for the ideal model. 
This results in $B_1$ becoming either ($\tilde{I}$) an almost cloaked region 
 that with a high probability does not accept energy $E$ particles from outside $\Sigma$;  or ($\tilde{II}$) a trap that supports {\it almost trapped} states, which correspond to a particle  at energy $E$ trapped in $B_1$ with high probability. 
A  design  of either type could possibly be implemented by an array of ac and dc electrodes 
with total effective potential approximating   $V_n^E$ for large $n$ \cite{MGW}.
This leads to a new type of trap for ions, differing from, e.g., 
the Paul  \cite{Paul},  Penning  \cite{Penn} or Zajfman  \cite{Zaj} traps,
and justified on the level of quantum mechanics. Furthermore, the trap may be made tunable by including a dc electrode in the trapped region, corresponding to a Coulomb $W$; varying the charge changes whether or not  $E$ is as  eigenvalue of $-\nabla^2 +W$ and thus which of ($\tilde{I}$) or ($\tilde{II}$) holds.

\emph{Topological ion traps.} The basic construction outlined above can be modified to make the wave function on $B_1$ behave as though it were confined to a compact, boundaryless three-dimensional manifold, topologically but not metrically the three-dimensional sphere, $\mathbb S^3$. By suitable choice of metric, the energy level $E$ can have arbitrary multiplicity for the interior of the resulting trap,    allowing one to implement physical systems mimicking  matter waves on abstract spaces.  As the starting point one uses not the original cloaking conductivity $\sigma_1$  (the {\it single coating} construction), but rather a {\it double} coating \cite[Sec.2]{GKLU1}, which we denote here by $\sigma^{(2)}$. This  is singular from both sides of $\Sigma$, and in the EM cloak setting corresponds to coating both sides of $\Sigma$ with metamaterials. 
See \cite[Fig. 7]{GKLU5}. By \cite[Sec. 3.3]{GKLU1}, the finite energy  solutions of the resulting Helmholtz equation  on $B_3$ split into direct sums of waves on $B_3-B_1$, as for $\sigma_1$, and waves on $B_1$ which are identifiable with eigenfunctions of the Laplace-Beltrami operator $-\nabla_g^2$ 
on $(\mathbb S^3,g)$ with eigenvalue $\sqrt{E}$. If one takes $g$ to be the standard metric on $\mathbb S^3$, then  nonground states are degenerate and of high multiplicity, while a generic choice of  $g$ yields simple energy levels \cite{Uhl}. By suitable choice of $g$,
any  finite number of  energy levels and multiplicities can be specified \cite{CdV}, allowing traps   supporting almost trapped states at energy $E$ of arbitrary degeneracy.

\emph{Magnetically tunable  quantum beam switch.} Consider a beam of ions of energy $E$, leaving an oven and traveling through a tube $T=\{0\le \rho\le \rho_0,\, 0\le\theta\le 2\pi,\, 0\le z\le L\}$ (in cylindrical coordinates). Treating the ions as matter waves, place in $T$ several almost trapping traps of the  type described above, centered at points on the $z$-axis,  forming  a potential $V=\sum_{j=1}^N V_n^E(z-z_j)$. The   techniques above may be applied to  magnetic  Schr\"odinger 
equations and DN operators on a region $\Omega$,
\vspace{-0.3cm}
\ba
\Big(-(\nabla+iA)^2 +V\Big)\psi = E\psi\quad\hbox{on }\Omega, \\
 \Lambda_{V,A}(E)(f)=  \p_\nu \psi|_{\p \Omega}+ i (A\cdot \nu) f\quad\hbox{on }\p \Omega.
\ea

We design the traps so that, in the absence of a magnetic field, or for small field strengths, the traps act as cloaks and thus the ions pass through $T$ unhindered. 
However,  if a homogeneous magnetic field is then applied to the tube, chosen so that the magnetic Schr\"odinger operator has $E$ as a Neumann eigenvalue inside each trap, then there is a large probability that an ion passing the $j$th trap will bind to that trap. If $N$ is large enough, then the probability that any ion traveling the length of $T$ will become bound is $\sim 1$, and $T$ thus functions as a magnetically controlled switch for the beam of ions.

\emph{Magnified magnetic fields.} 
For a homogeneous magnetic field with linear magnetic potential $A$, one can obtain a sequence of electrostatic potentials  $V_n$
for which $\lim_{n\to\infty} \Lambda_{V_n,A}(E) =\Lambda_{0,\tilde{A}}(E)$, with $\tilde{A}$ singular at a point. I.e., in the presence of a homogeneous magnetic field, the $V_n$   produce scattering or boundary measurements  that tend, as $n\to\infty$, to those of the zero electrostatic potential in the presence of a magnetic field blowing up at a point, 
$|\tilde{A}(\rx)|\sim r^{-1},\, |\tilde{\mathbf B}(\rx)|\sim r^{-2}$, giving the illusion of a locally singular magnetic \nolinebreak field \cite{GKLU5}.

{\bf Acknowledgements:} AG and GU
are supported by NSF, GU by a Walker Family Professorship, ML by Academy of Finland and YK by EPSRC.
We are grateful for 
discussions  with A. Cherkaev and V. Smyshlyaev on homogenization and  S. Siltanen on numerics.

\bibliographystyle{amsalpha}

\end{document}